\documentstyle[12pt]{article}
\def\ltsim{\raise 2pt \hbox {$<$} \kern-1.1em \lower 4pt \hbox {$\sim$}}
\def\gtsim{\raise 2pt \hbox {$>$} \kern-1.1em \lower 4pt \hbox {$\sim$}}
\begin{document}
\input psfig

\vspace{1.25cm}
\Large

{\bf
%
SPACE VLBI OBSERVATIONS OF MKN 501
}
\normalsize

\vspace{0.5cm}

\newcommand{\au}[2]{#1$^{#2}$}
%
\au{G. Giovannini}{1,2}, \au{W.D. Cotton}{3}, \au{L. Feretti}{2}, 
\au{L. Lara}{4}, and \au{T.Venturi}{2}

\vspace{0.5cm}

\newcommand{\ins}[2]{$^{#1}$ {\it #2}\\}
%
\ins{1}{Dipartimento di Fisica, Universit\'a di Bologna, via B. Pichat 6/2, 40126 Bologna, Italy}
\ins{2}{Istituto di Radioastronomia del CNR, Via P. Gobetti 101, I-40129
Bologna, Italy}
\ins{3}{National Radio Astronomy Observatory, 520 Edgemont Rd, Charlottesville
VA 22903-2475, USA}
\ins{4}{Instituto de Astrof\'{\i}sica de Andaluc\'{\i}a, CSIC, Apdo. 3004,
18080 Granada, Spain}
\vspace{0.5cm}

ABSTRACT

%
We present here two epochs of Space VLBI Observations at 18 cm of the BL-Lac
type object Mkn 501. Thanks to the high resolution of these new data we have
found that the inner jet is centrally brightened at its beginning but becomes
extended and limb brightened at $\sim$ 8 mas from the core. Moreover a comparison
between the two epochs shows the presence of a proper motion with apparent
velocity = 6.7c. Observational data have been used to constrain the jet
velocity and orientation.

\vspace{0.5cm}

INTRODUCTION AND OBSERVATIONS

Markarian 501 (Mkn 501) is a BL-Lac type object at z = 0.034. It is a well
studied source in radio, optical and X-ray bands being one of the brightest
Bl-Lac object at all wavelengths. It was the second source
(after Mkn 421) to be detected at TeV energies.
In the radio band it has a total radio power at 408 MHz of 8.9 $\times 10^{24}$
W/Hz consistent with the expectation of unified scheme models that BL-Lac type
sources are FR I galaxies seen on the line of sight (we use H$_0$ = 50 km sec$^{-1}$ Mpc$^{-1}$).

We observed this source
in August 4, 1997 during an orbit checkout of the HALCA satellite, with the
following observing array:
VLBA (observing time 7$^h$); Goldstone (4$^h$); HALCA with the Green Bank
tracking station (3$^h$ only). The data were correlated in Socorro,
calibrated and reduced with the AIPS package (15 Apr. 1997 version).
Only the 1st IF gave useful data.
On April 8, 1998 the source was observed again for 10$^h$ with
the following array: VLBA, Goldstone, Robledo, HALCA with the Green Bank and
Madrid tracking stations. The data were correlated in Socorro and reduced
with the AIPS package (15 Apr 1998 version). We found good data from all the
telescopes and for both IFs. Amplitude calibration was done initially for
VLBA telescopes only and after applied to the whole array. All data were
globally fringe fitted and then self-calibrated.

\vspace{0.5cm}

SOURCE MORPHOLOGY AND JET DYNAMICS

At parsec resolution Mkn 501 shows a one-sided jet structure which changes
its orientation from $\sim$ 140$^\circ$ to $\sim$ 30$^\circ$.
In Fig. 1 we show a full resolution map obtained from
uniform weighted data, 
while in Fig. 2 a map obtained with a larger beam is shown. 
\begin{figure}
\centerline{\psfig{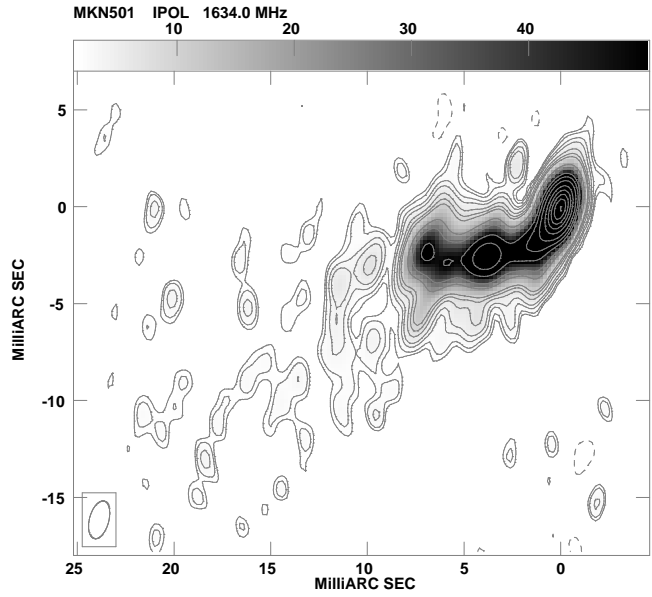}}
\caption{\it Full resolution Mkn 501 map obtained with space VLBI. The HPBW
is 2 $\times$ 1 mas at PA = -15$^\circ$. The noise level is 0.7 mJy/beam.
Levels are: -2 1.5 2 3 4 6 8 10 20 30 50 70 100 150 200 250 300 350 mJy/beam.
}
\label{figopt1}
\end{figure}
\begin{figure}
\centerline{\psfig{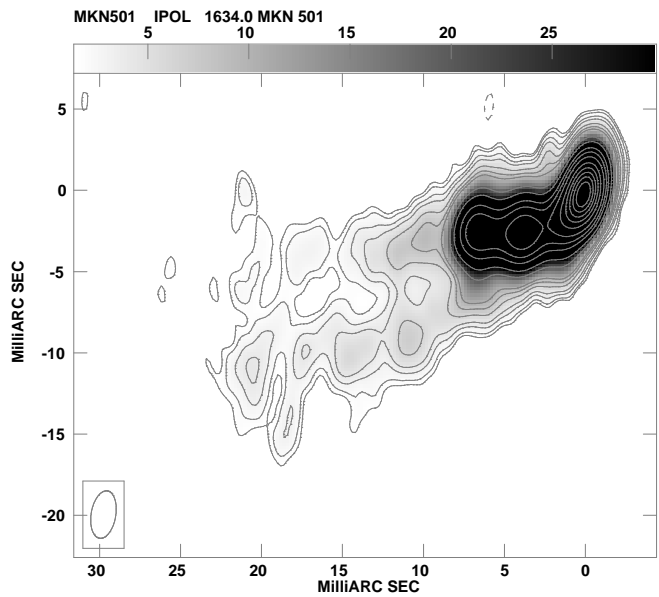}}
\caption{\it 
Space VLBI map of Mkn 501 at the resolution of 2.9 $\times$ 1.5
mas (PA = -10 $^\circ$). The noise level is 0.3 mJy/beam. Levels are: -2 1.5
2 3 4 6 8 10 20 30 50 70 100 150 200 250 300 350 mJy/beam.
}
\label{figopt1}
\end{figure}
\begin{figure}
\centerline{\psfig{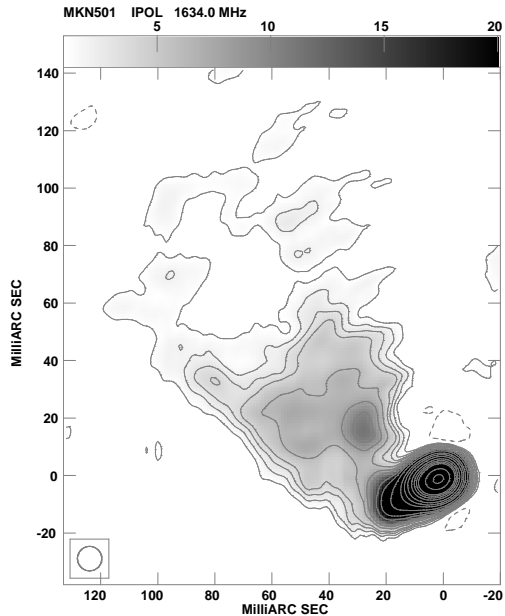}}
\caption{\it 
VLBI map of Mkn 501 with only ground based telescopes.
The HPBW is 8.5 mas and
the noise level is 0.35 mJy/beam. Levels are: -2 1
2 3 4 6 8 10 20 30 50 70 100 150 200 250 300 500 700 mJy/beam.
}
\label{figopt1}
\end{figure}
From an analysis of the two maps we note a clear 
change in the jet structure at about 8 mas from the core.  At the beginning 
the jet is transversally resolved but it is centrally brightened. 
After an elongated feature the jet brightness becomes limb brightened: the 
maximum of
the surface brightness is clearly on both sides of the parsec scale jet.
We interpret this observational result as an indication of a change
in the jet physical properties at $\sim$ 8 mas (7.4 pc) from the core. 
The limb brigthened jet is visible for 20-25 mas
($\sim$ 20 pc) after where it shows a large change in its Position Angle (PA)
and a dramatic expansion (Fig. 3). This extented jet is visible for more than 
100 mas and appears edge brightened. No evidence of a helical
structure is visible in our maps.

\vspace{0.5cm}

JET ORIENTATION AND VELOCITY

We produced identical maps from the first and second epoch data with a HPBW =
2.9 $\times$ 1.5 mas in PA -9.7$^\circ$. The two maps are in good agreement, 
but suggest the existence of proper motion between the two epochs. We are aware that
to measure
a proper motion with only two epochs data may give unreliable results, however
the good quality of present data and the agreement between the whole structure
strongly suggests the presence of a proper motion with apparent velocity $\sim$
6.7c. More Space VLBI Observations have been requested to confirm this result.
We note that in the second epoch, the core is resolved and with a lower
peak flux density if compared with the first epoch, suggesting that a new 
component is
emerging out. Such an apparent velocity implies that the real jet velocity has 
to be \gtsim 0.989c and the jet has to be oriented at an angle smaller than
17$^\circ$.

We used the available VLBI data to derive constraints on the jet velocity and
orientation (see Giovannini et al. 1994, ApJ {\bf 435}, 116) for a more detailed discussion).
Assuming that parsec scale jets are intrinsically two-sided and symmetric, we
can constrain the jet velocity and orientation from the observed jet asymmetry.
In our maps the Jet/CounterJet ratio is \gtsim 200 and therefore we can give
the following constraint on the jet velocity ($\beta$ = v/c) and orientation
($\theta$): $\beta$ cos$\theta$ \gtsim 0.79.
Moreover from the known correlation between the core power and the unbeamed total
radio power (Giovannini et al., 1988, A.A. {\bf 199}, 73) we can derive the expected
intrinsic core radio power from the observed total radio power. 
Comparing the expected and the observed core radio power
we obtain that Mkn 501 has to be oriented at $\theta$ \ltsim 26$^\circ$ with 
$\beta$ in the range 1 - 0.88.

If we assume that the bulk and pattern jet velocity are comparable in the
inner parsec scale jet, we can compare the constraints derived from the J/CJ ratio and the core dominance with those obtained from the measured proper motion.
From these data we derive that Mkn 501 is oriented at
 $\sim$ 10$^\circ$ - 15$^\circ$ with respect to the line of sight, with a velocity in the range 0.990 - 0.999 c.
We note that a higher velocity or a smaller angle implies a too high doppler
factor which would give a too strong parsec radio core with respect to
the total flux density.

\vspace{0.5cm}

DISCUSSION AND CONCLUSIONS

Thanks to the high resolution provided by Space VLBI Data we have observed 
that the parsec scale
jet of Mkn 501 shows a change in its physical properties at 8 mas from the
core: the inner jet is centrally brightened, but after a vertical shock-like
structure it becomes limb brightened. The change in the morphological shape 
reflects a change in the jet physical properties probably due to its interaction with the surrounding medium. Moreover comparing the two epoch maps available
we have measured a possible proper motion with an apparent velocity 
$\sim$ 6.7 c.
From VLBI data we have derived that Mkn 501 is oriented at an angle of 
10$^\circ$ - 15$^\circ$ and has an intrinsic velocity in the range 0.990 - 
0.999 c. We note that current models of gamma ray emission and intra-day 
variability suggest that in the inner region
(0.003 - 0.03 pc) the jet should be oriented at $\theta$ \ltsim 5.7 $^\circ$
and should have a Lorentz factor $\gamma$ \gtsim 10 (see e.g. Salvati et al., 
1998, ApJ in press). These figures seem to be in contrast with the derived 
constraints from VLBI radio data, however we note that the regions
where the radio emission ({\it the core region}) and the gamma and X-Ray emission
are produced, are not coincident, being the former at a larger distance from the core
than the latter.
We estimate that the {\it radio core} has a size \gtsim 0.05 pc. 
Therefore radio and gamma ray data are in agreement if when moving from
0.003 pc to 0.05 pc the jet orientation changes from $\sim$ 5$^\circ$ to $\sim$
10$^\circ$. Such a change in the jet orientation is consistent 
with the distorted
morphology found in the parsec scale of Mkn 501. A jet velocity decrease is not
necessary if the measured apparent velocity is confirmed by new data. However
morphological changes in the jet suggests that jet deceleration possibly 
exists in the transition region discussed above.  

\underline{Acknowledgments.}  

We gratefully acknowledge the VSOP Project, which is led by the Japanese
Institute of Space and Astronautical Science in cooperation with many
organizations and radio telescopes around the world. The NRAO is operated by
Associated Universities, Inc. under a cooperative agreement with the National 
Science Foundation.
\end{document}